\documentclass[conference,9pt]{IEEEtran}
\IEEEoverridecommandlockouts
\usepackage[T1]{fontenc}
\usepackage{amsmath,amssymb}
\usepackage{graphicx,booktabs,array}
\usepackage{xcolor}
\usepackage{tikz}
\usetikzlibrary{arrows.meta,positioning}
\usepackage{algorithm}
\usepackage{algpseudocode}
\usepackage{float}
\usepackage[checkfloat]{flushend}
\usepackage{cite}
\usepackage{microtype}
\usepackage{url}
\usepackage[hidelinks]{hyperref}
\hypersetup{pdftitle={Structure-Guided Synthesis of Low-Error Syndrome Extraction Circuits},
  pdfauthor={Shanchuan Li, Xin Hong, Pengcheng Zhu, Xiangzhen Zhou, Fanxu Meng, Shenggang Ying},pdfsubject={Research preprint}}
\newcommand{\F}{\mathbb{F}_2}
\newcommand{\ket}[1]{\lvert #1\rangle}
\newcommand{\cx}[1]{\mathsf{X}_{#1}}
\newcommand{\cz}[1]{\mathsf{Z}_{#1}}
\newcommand{\ind}[1]{\mathbf{1}\!\left[#1\right]}
\newcommand{\dcirc}{d_{\mathrm{circ}}}
\newcommand{\LER}{\operatorname{LER}}

\definecolor{orderblue}{RGB}{32,112,145}
\definecolor{faultred}{RGB}{178,66,44}
\definecolor{ordergreen}{RGB}{40,126,113}
\graphicspath{{figures/}}

\usepackage[normalem]{ulem}

\title{Structure-Guided Synthesis of Low-Error\\Syndrome Extraction Circuits}
\author{
\IEEEauthorblockN{Shanchuan Li}
\IEEEauthorblockA{\textit{Arclight Quantum Computing Inc.}\\
Beijing, China\\
lishanchuan217@gmail.com}
\and
\IEEEauthorblockN{Xin Hong}
\IEEEauthorblockA{\textit{Institute of Software,}\\
\textit{Chinese Academy of Sciences}\\
Beijing, China\\
hongxin@ios.ac.cn}
\and
\IEEEauthorblockN{Pengcheng Zhu}
\IEEEauthorblockA{\textit{Taizhou University}\\
Taizhou, Jiangsu, China\\
zhupcnt@163.com}
\and[\hfill\mbox{}\par\vspace{1em}\mbox{}\hfill]
\IEEEauthorblockN{Xiangzhen Zhou\textsuperscript{*}}
\IEEEauthorblockA{\textit{Arclight Quantum Computing Inc.}\\
Beijing, China\\
\textit{Nanjing Tech University}\\
Nanjing, China\\
zhouxiangz@qq.com}
\and
\IEEEauthorblockN{Fanxu Meng\textsuperscript{*}}
\IEEEauthorblockA{\textit{Nanjing Tech University}\\
Nanjing, China\\
mengfanxuchina@163.com}
\and
\IEEEauthorblockN{Shenggang Ying}
\IEEEauthorblockA{\textit{Institute of Software,}\\
\textit{Chinese Academy of Sciences}\\
Beijing, China\\
yingsg@ios.ac.cn}
\thanks{\textsuperscript{*}Corresponding authors: Xiangzhen Zhou and Fanxu Meng.}
}

\begin{document}
\flushbottom
\maketitle

\begin{abstract}
Gate orders can change fault propagation and logical error rates
even when syndrome-extraction circuits implement the same ideal
stabilizer measurements.
Optimizing these orderings independently for every check,
however, becomes computationally demanding as code size increases.
We present a synthesis method for structured Calderbank--Shor--Steane
codes that groups checks by recurring connectivity and shares one
relative gate order per group, reducing independent ordering choices.
A CP-SAT solver jointly determines the shared orders and the
execution time of each physical gate, enforcing resource and
measurement constraints across the circuit.
Noisy circuit simulations with decoding then guide the selection
of the final circuit.
We compare with state-of-the-art approaches, including
Auto-Stabilizer-Check (ASC), PropHunt, and AlphaSyndrome.
Across 15 quantum low-density parity-check codes from ten code families, our method
achieves the lowest logical error rate estimates. Relative to ASC, the strongest
baseline in our evaluation, the median reduction is 11.4\% and the
largest is 38.6\%.
Our circuits improve circuit-level distance
over ASC on ten instances and preserve the code distance in seven.
Across five timed codes, our method reduces the mean time to
the first feasible circuit by factors of 21.4--136.6 relative to ASC,
at equal or lower depth.

\end{abstract}


\section{Introduction}
\label{sec:intro}

Quantum error correction is essential for reliable quantum
computation in the presence of physical noise.
Stabilizer codes detect errors through repeated measurements of
commuting checks that preserve logical information~\cite{terhal2015quantum}.
Within this framework, quantum low-density parity-check
(qLDPC) codes use sparse checks and offer a promising route
to reducing physical-qubit
overhead~\cite{breuckmann2021quantum,bravyi2024high}.
Realizing these advantages requires syndrome-extraction circuits
that preserve the code's protection under physical noise.
This code-to-circuit link also motivates recent code-discovery workflows
that use circuit-level evaluation to assess candidate codes~\cite{yan2026omniqec}.

Syndrome extraction can itself introduce correlated data errors.
A fault on an ancilla can spread through subsequent
controlled-NOT (CNOT) gates, producing a hook
error~\cite{dennis2002topological}.
The code specifies the required interactions, leaving their
order and timing as design choices.
Gate order determines which data qubits a fault reaches,
while gate timing influences idle periods and the interleaving
of check measurements.
Circuits that measure the same checks at the same depth can
consequently have different logical error rates (LERs).
Even low-weight checks admit many orders. Shared data qubits
and X/Z interleaving constraints couple these choices: an order
for one check can block another's schedule.

Existing methods address this search in several ways.
Auto-Stabilizer-Check (ASC) uses circuit-validity constraints
to find low-depth schedules and selects candidates by
LER~\cite{zhang2026optimal}.
AlphaSyndrome uses decoder feedback to guide a Monte Carlo tree
search~\cite{liu2026alphasyndrome}, while PropHunt uses
fault-propagation analysis to guide circuit
modifications~\cite{viszlai2026prophunt}.
Structure also guides code-specific designs. Bravyi et al.~\cite{bravyi2024high}
construct seven-layer CNOT schedules for BB codes.
In their schedules, corresponding gates occupy the same layer across
checks of each type.
For automated synthesis across code constructions, the challenge is to
express such ordering relationships as group-level decisions and connect
them to the established scheduling constraints.

We introduce a structure-guided synthesis approach that enforces shared relative CNOT orderings among checks with matching connectivity. This cuts down the number of separate ordering choices, and a single shared ordering determines the associated hook-error patterns for every check within a group. The approach automatically identifies sharing groups and port mappings from the check matrices and the construction representation. The combined model ties each group’s ordering choices to the per-gate execution times of its CNOTs. CP-SAT~\cite{perron2023cp} jointly optimizes both aspects while respecting resource and measurement constraints across the circuit. Equivalent gates may be placed in different layers to prevent conflicts. Circuit simulation and decoding then choose a schedule with low LER.

We compare our method with state-of-the-art approaches to
syndrome-extraction circuit synthesis, including ASC, PropHunt,
and AlphaSyndrome. The evaluation covers 15 structured CSS qLDPC codes
from ten code families
under a common noise model and decoder.
ASC is the strongest baseline on all 15 codes.
Our circuits have lower LER estimates on every code, with a median reduction of
11.4\% and a maximum of 38.6\%.
The advantage also holds in all thirteen equal-depth comparisons.
Our circuits improve circuit-level distance over ASC on ten
instances and preserve the code distance in seven.
Sharing orders lowers mean LER by 13.4\%--46.6\% in an ablation.
Across five timed codes, our method reduces the mean time to the
first feasible circuit by factors of 21.4--136.6 relative to ASC.
The resulting depths are equal or lower.
\section{Background and Motivating Example}
\label{sec:background}

\subsection{CSS Codes and Measurement Circuits}

A stabilizer code detects errors by measuring commuting Pauli
operators called \emph{checks}, while preserving the encoded
logical information. Encoded states have eigenvalue $+1$ for every
check. We encode the $+1/-1$ outcomes as $0/1$ to form the
\emph{syndrome}. A decoder uses syndromes across rounds to infer errors, accounting
for data and measurement faults. Recent neural approaches address
inference latency and the cost of training decoders for larger
codes~\cite{yan2026rethink,yan2026efficient}.

Calderbank--Shor--Steane (CSS) codes have two types of checks:
products of Pauli $X$ operators and products of Pauli $Z$
operators~\cite{calderbank1996good,steane1996multiple}.

\begin{samepage}
We use the standard notation
$\lbrack\!\lbrack n,k,d\rbrack\!\rbrack$:
the code stores $k$ logical qubits in $n$ data qubits
and has \emph{code distance} $d$.
The distance is the minimum weight of a Pauli operator that changes
logical information while preserving all check outcomes.
Its weight counts the data qubits on which it acts nontrivially.

\end{samepage}

Each check uses one auxiliary qubit, or \emph{ancilla}.
To measure an X check, we prepare its ancilla in $\ket{+}$.
We apply a CNOT from the ancilla to each data qubit in the check,
then measure the ancilla in the X basis.
To measure a Z check, we prepare the ancilla in $\ket{0}$,
use it as the target of CNOTs controlled by the data qubits,
and measure it in the Z basis.

We use a twelve-qubit code as a running example, called the $C_6$ example.
Its data qubits are $L_0,\ldots,L_5,R_0,\ldots,R_5$,
and its checks are $\cx{u}$ and $\cz{u}$ for $u\in\{0,\ldots,5\}$.
The \emph{support} of a check is the set of data qubits on which it acts.
In this example, the supports are
\begin{equation}
\begin{aligned}
 Q_{\cx{u}}&=\{L_u,L_{u+2},R_u,R_{u+1}\},\\
 Q_{\cz{u}}&=\{L_u,L_{u-1},R_u,R_{u-2}\},
\end{aligned}
\label{eq:example}
\end{equation}
where all subscripts are taken modulo six. For example, $\cx{0}$
measures $X_{L_0}X_{L_2}X_{R_0}X_{R_1}$, where $X_q$ denotes Pauli
$X$ on qubit $q$.
Each X check overlaps each Z check on an even number of qubits,
so the checks commute.
Translations of all subscripts modulo six form the cyclic group $C_6$
and preserve the connectivity between checks and data qubits.
Adding two to every subscript maps $\cx{0}$ to $\cx{2}$.

The $C_6$ example has the polynomial description $a(x)=1+x^2$
and $b(x)=1+x$, where $x$ shifts a subscript forward by one modulo
six. X checks use $a(x)$ on the L qubits and $b(x)$ on the R qubits;
Z checks exchange the two polynomials and reverse the shifts.
These polynomials specify the repeated connectivity of the checks.

For a general CSS code, binary parity-check matrices
$H_X\in\F^{m_X\times n}$ and $H_Z\in\F^{m_Z\times n}$ record
the supports of $m_X$ X checks and $m_Z$ Z checks. Rows represent
checks, columns represent data
qubits, and ones indicate interactions. The checks commute exactly when
\begin{equation}
 H_XH_Z^{T}\equiv 0\pmod{2},
 \label{eq:css}
\end{equation}
where all operations are over the binary field $\F$.
A check's \emph{weight} is the number of data qubits in its support.
A code family is qLDPC when its row and column weights remain
bounded independently of code size.
Expanding a code's construction representation gives the physical
interactions recorded in the rows of $H_X$ and $H_Z$.

\subsection{Why Gate Order Matters}

A CNOT propagates an X error from its control to its target
and a Z error from its target to its control.
An X error on an X-check ancilla, or a Z error on a Z-check ancilla,
can therefore spread to several data qubits through the remaining CNOTs.
The resulting correlated data error is often called a
\emph{hook error}~\cite{dennis2002topological}.
Which qubits it affects depends on which CNOTs remain after the fault occurs.

\begin{figure}[!tbp]
 \centering
\begin{tikzpicture}[x=.92cm,y=.48cm,font=\footnotesize,
  control/.style={circle,fill=black,inner sep=1.35pt},
  fault/.style={draw=faultred,fill=white,text=faultred,inner sep=1.5pt}]
\foreach \yy/\lab in {0/{a:\ket{+}},-1/{L_0},-2/{R_0},-3/{R_1},-4/{L_2}} {
  \node[anchor=east] at (.75,\yy) {$\lab$};
  \draw (.85,\yy)--(6.85,\yy);
}
\node[anchor=west] at (.05,1.05) {(a) Order: $L_0,R_0,R_1,L_2$};
\foreach \xx/\yy in {1.4/-1,2.65/-2,3.9/-3,5.15/-4} {
  \draw (\xx,0)--(\xx,\yy);
  \node[control] at (\xx,0) {};
  \draw[fill=white] (\xx,\yy) circle[x radius=.10cm,y radius=.10cm];
  \draw (\xx-.105,\yy)--(\xx+.105,\yy);
  \draw (\xx,\yy-.205)--(\xx,\yy+.205);
}
\node[fault] at (3.25,0) {$X$};
\node[draw,fill=white,inner sep=2pt] at (6.35,0) {$M_X$};
\node[fault] at (6.35,-3) {$X$};
\node[fault] at (6.35,-4) {$X$};
\begin{scope}[yshift=-3.25cm]
\foreach \yy/\lab in {0/{a:\ket{+}},-1/{L_0},-2/{R_0},-3/{R_1},-4/{L_2}} {
  \node[anchor=east] at (.75,\yy) {$\lab$};
  \draw (.85,\yy)--(6.85,\yy);
}
\node[anchor=west] at (.05,1.05) {(b) Order: $L_0,R_1,R_0,L_2$};
\foreach \xx/\yy in {1.4/-1,2.65/-3,3.9/-2,5.15/-4} {
  \draw (\xx,0)--(\xx,\yy);
  \node[control] at (\xx,0) {};
  \draw[fill=white] (\xx,\yy) circle[x radius=.10cm,y radius=.10cm];
  \draw (\xx-.105,\yy)--(\xx+.105,\yy);
  \draw (\xx,\yy-.205)--(\xx,\yy+.205);
}
\node[fault] at (4.5,0) {$X$};
\node[draw,fill=white,inner sep=2pt] at (6.35,0) {$M_X$};
\node[fault] at (6.35,-4) {$X$};
\end{scope}
\end{tikzpicture}

 \caption{\textbf{Fault propagation for two orders of check $\cx{0}$.}
 Red $X$ symbols mark the ancilla fault after the CNOT with
 $R_0$ and the resulting data errors.}
 \label{fig:fault}
\end{figure}
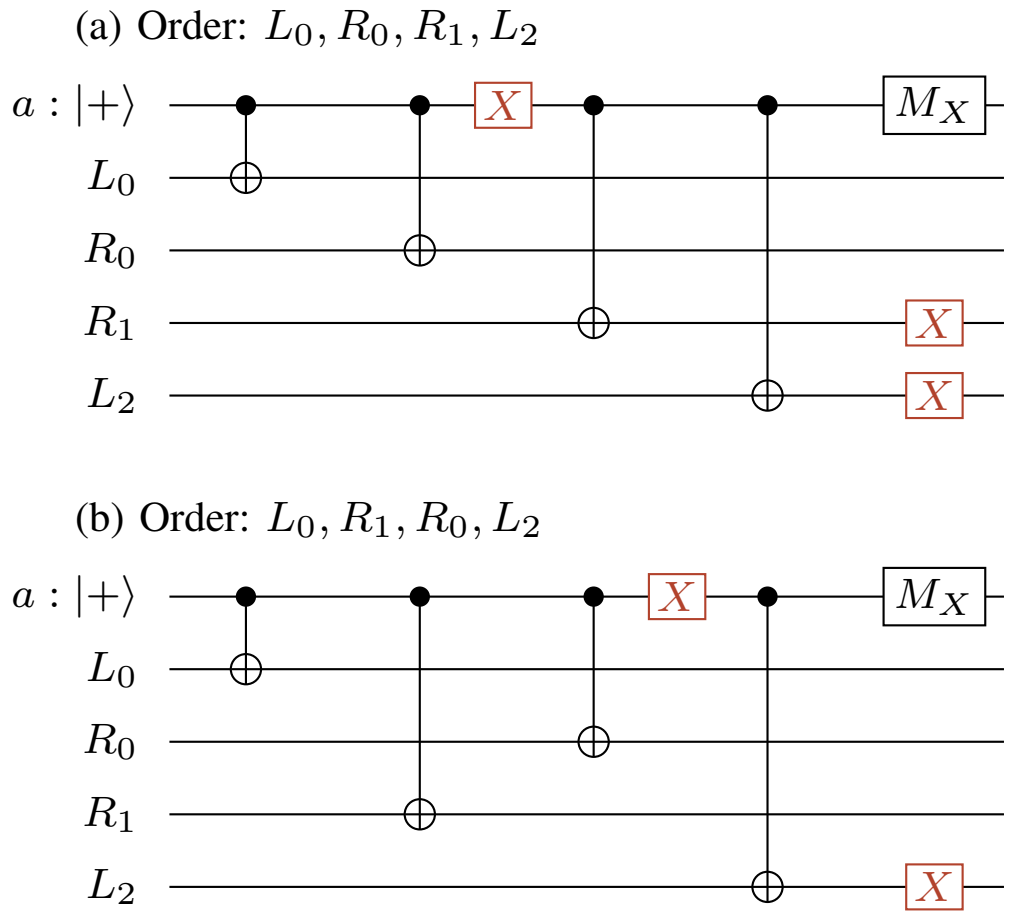

Fig.~\ref{fig:fault} illustrates this dependence using $\cx{0}$.
In panel (a), the ancilla controls CNOTs targeting the data qubits
in the order $(L_0,R_0,R_1,L_2)$.
An X fault immediately after the CNOT with $R_0$ spreads through
the final two CNOTs, producing the data error $X_{R_1}X_{L_2}$.
Panel (b) exchanges the CNOTs targeting $R_0$ and $R_1$.
The fault after the CNOT with $R_0$ then spreads only to $L_2$,
because only that CNOT remains.

\subsection{Circuit Validity and Performance Metrics}

A \emph{schedule} assigns each required CNOT to an execution layer,
with all gates in a layer running in parallel.
Its CNOT depth is the number of layers, excluding ancilla preparation
and readout.
To avoid resource conflicts, gates sharing a qubit must run
in different layers.
The schedule must also interleave the CNOTs of X and Z checks
so that the circuit measures the intended checks.
Although the checks commute, arbitrary CNOT interleavings
can change the measurement~\cite{tremblay2022constant,zhang2026optimal}.

The \emph{circuit-level distance} counts physical faults, each of
which may affect several data qubits. For schedule $\sigma$, fix
the measurement task, boundary conditions, and fault model. Let
$\mathcal{E}_{\sigma}$ be the allowed fault configurations and
$|f|$ the number of faults in $f$. A \emph{detector} is a measurement
parity with a known noiseless value. The binary vectors
$s_{\sigma}(f)$ and $\ell_{\sigma}(f)$ record flips of detectors
and tracked logical observables, respectively. The distance is
\begin{equation}
 \dcirc(\sigma)=\min_{f\in\mathcal{E}_{\sigma}}
 \bigl\{|f|:s_{\sigma}(f)=0,\ \ell_{\sigma}(f)\ne0\bigr\}.
 \label{eq:circuit-distance}
\end{equation}
Equation~\eqref{eq:circuit-distance} counts the fewest faults
that flip a logical observable without triggering a detector.

Manes and Claes~\cite{manes2025distance} establish distance robustness
for HGP codes. Under their stabilizer-measurement model, HGP codes
retain their effective distance for any valid CNOT ordering.

The \emph{logical error rate} (LER) measures how often decoding fails
under a specified noise model.
X- and Z-basis memory experiments evaluate LER by preparing known
logical states, repeating extraction rounds, and comparing the
final readout with the decoder's predictions.
For memory basis $b\in\{X,Z\}$, let $F_b$ be the probability of
decoding failure after $R$ rounds.
We first normalize to an effective failure probability per logical
qubit, $q_b=1-(1-F_b)^{1/k}$. For $q_b\leq 1/2$, the per-round
normalization~\cite{bausch2024learning} gives
\begin{equation}
 r_b=\frac{1-(1-2q_b)^{1/R}}{2},\qquad
 \LER=\frac{r_X+r_Z}{2}.
 \label{eq:ler}
\end{equation}
Thus, LER denotes the effective error rate per logical qubit per
round, averaged over the two memory bases.

\section{Problem Formulation}
\label{sec:problem}

The input check matrices $H_X,H_Z$ fix the required ancilla--data
interactions. We seek gate orders and execution times that implement
these interactions as a valid, low-LER syndrome-extraction circuit.
The accompanying construction representation supplies the
correspondences between checks and their interactions, as illustrated
by the $C_6$ example.

\emph{Design variables.}
Let $m=m_X+m_Z$ be the total number of checks,
indexed by $i=1,\ldots,m$.
Check $i$ has support $Q_i$ and weight $w_i=|Q_i|$.
With depth limit $D$, each required CNOT has an integer time
\begin{equation}
 t_{i,q}\in\{1,\ldots,D\}
 \label{eq:time}
\end{equation}
where $t_{i,q}$ is the layer in which the CNOT between check $i$'s
ancilla and data qubit $q\in Q_i$ executes.
We denote the collection of these times by $\sigma$.
These times determine gate order within each check and each qubit's idle periods.

\emph{Feasible schedules.}
Let $\mathcal{F}_D(H_X,H_Z)$ contain all schedules within depth
bound $D$ that implement the required joint check measurements.
Each required CNOT executes exactly once, and gates sharing an
ancilla or data qubit occupy distinct layers.
The interleaving of X- and Z-check CNOTs must preserve the intended
measurements.

\emph{Performance objective.}
For fixed noise, decoding, and memory-experiment settings, the
design objective is
\begin{equation}
 \min_{\sigma\in\mathcal{F}_D(H_X,H_Z)}\LER(\sigma),
 \label{eq:objective}
\end{equation}
where $\LER(\sigma)$ is the effective rate defined
in~\eqref{eq:ler} for the circuit implementing $\sigma$.
The depth bound applies to each extraction round.
We output schedule $\sigma^\star$ and its LER estimate.

Solving~\eqref{eq:objective} requires coordinating local orders across
the circuit, since an order that works for one check may conflict with
the gates or measurements of others. These coupled choices span
$\prod_{i=1}^{m}w_i!$ combinations before circuit constraints are applied,
as each weight-$w_i$ check admits $w_i!$ local orders.
Evaluating a candidate's LER further requires noisy simulation and decoding.

\section{Structure-Guided Circuit Synthesis}
\label{sec:method}

Construction index maps define sharing groups and port maps.
Using these maps, a joint model chooses shared orders and execution
times. Circuit simulation and decoding then select the final schedule.

\subsection{Structure-Guided Shared Ordering}
\label{sec:templates}

Checks with corresponding connectivity share one relative CNOT order,
so a single ordering choice determines the local CNOT sequence of
every check in a group.
The code's construction representation supplies index maps between
checks with corresponding connectivity. Each map pairs the data-qubit
indices in their supports, matching the nonzero entries of their
check-matrix rows. The algorithm groups checks related by these maps,
with X and Z checks treated separately.
It labels the interactions of a representative check as ports and
carries these labels to the other members through the index maps.
The groups therefore determine which checks share an order,
while the port maps translate that order into CNOTs on specific data qubits.

For the bicycle-type and Haah instances, periodic-coordinate
translations define the correspondences. For abelian lifted-product
and fiber-bundle codes, copies of the same base check along the lift
or fiber coordinate share an order, with ports identified by data block
and offset. The non-Abelian two-block group-algebra instances use
base-check types and construction-term labels to define shared ports.
For the hypergraph-product and random quantum Tanner instances,
the construction representation includes verified factor or
Tanner-graph automorphisms, from which the algorithm derives
sharing groups and port maps.

Let $\mathcal{G}$ denote the collection of sharing groups.
Checks in each group $C\in\mathcal{G}$ have a common weight $w_C$.
For check $i$, the port map is a bijection
\begin{equation}
 \phi_i:\{1,\ldots,w_i\}\longrightarrow Q_i,
 \label{eq:port}
\end{equation}
where $\phi_i(p)$ is the data qubit involved in the CNOT labeled $p$.
The label identifies a corresponding construction role across member
checks. In particular, construction-term labels provide these
correspondences even for codes without global translation symmetry.
The joint model expresses sharing through $\mathcal{G}$ and
$\{\phi_i\}$, using the same ordering and scheduling constraints
for each construction.

In the $C_6$ example, translations modulo six define two sharing
groups: all six X checks and all six Z checks.
The same translations establish the port maps
shown in Table~\ref{tab:example}.
For example, port 2 of $\cx{0}$ labels the CNOT involving $L_2$.
Translating all indices by two maps it to the CNOT involving $L_4$
in $\cx{2}$, where it retains label 2.

\begin{table}[!tbp]
 \caption{Port maps and a valid four-layer CNOT schedule for $C_6$}
 \label{tab:example}
 \centering
 \small
 \setlength{\tabcolsep}{3.5pt}
 \begin{tabular}{@{}lcccc@{\hspace{10pt}}cccc@{}}
 \toprule
 & \multicolumn{4}{c}{Data qubit by port}
 & \multicolumn{4}{c}{Data qubit by layer}\\
 \cmidrule(lr){2-5}\cmidrule(l){6-9}
 Check & 1 & 2 & 3 & 4 & 1 & 2 & 3 & 4\\
 \midrule
 $\cx{0}$ & $L_0$ & $L_2$ & $R_0$ & $R_1$
          & $L_0$ & $R_0$ & $R_1$ & $L_2$\\
 $\cx{2}$ & $L_2$ & $L_4$ & $R_2$ & $R_3$
          & $L_2$ & $R_2$ & $R_3$ & $L_4$\\
 \midrule
 $\cz{0}$ & $L_0$ & $L_5$ & $R_0$ & $R_4$
          & $R_4$ & $L_0$ & $L_5$ & $R_0$\\
 \bottomrule
 \end{tabular}
\end{table}

Each group $C$ has a variable permutation $\pi_C$
representing its shared gate order.
Check $i\in C$ applies its CNOTs in the data-qubit sequence
\begin{equation}
 \phi_i\bigl(\pi_C(1)\bigr),\ldots,
 \phi_i\bigl(\pi_C(w_C)\bigr),\qquad i\in C.
 \label{eq:sequence}
\end{equation}
The sharing groups $\mathcal{G}$ and port maps $\{\phi_i\}$ are
derived from the construction representation and remain fixed during
candidate generation, whereas the model jointly chooses the shared
orders $\{\pi_C\}$ and execution times $\{t_{i,q}\}$.
Table~\ref{tab:example} uses $\pi_X=(1,3,4,2)$ for six
X checks and $\pi_Z=(4,1,2,3)$ for six Z checks.

A propagating ancilla fault at a given point in the shared order
reaches the same remaining ports within each member check.
Choosing one order therefore fixes the corresponding hook-error
supports for all checks in the group.

Before circuit constraints are imposed, sharing changes the
number of ordering combinations from $\prod_i w_i!$ to
$\prod_{C\in\mathcal{G}}w_C!$.
For $C_6$, twelve independent permutations become two shared
permutations, reducing the count from $(4!)^{12}$ to $(4!)^2$.
Meanwhile, all 48 CNOT execution times remain variables in the joint model.

We impose sharing to focus the available construction and evaluation
budgets on ordering patterns coordinated by the code's construction.
At a fixed depth, the unshared model contains every shared-order
schedule, but reaching these schedules depends on the solver's search.
By tying the ordering decisions, sharing changes which candidates the
solver finds within those budgets.
The ablation in Section~\ref{sec:ablation} compares shared and unshared
models in the same implementation to assess the effect on circuit quality.

\subsection{Joint Synthesis of Orders and Times}
\label{sec:constraints}

The feasibility of shared orders depends on resource
and measurement constraints across the entire circuit.
For example, at $D=4$ in the $C_6$ example, keep the X order
$(1,3,4,2)$ and change the Z order to $(1,4,2,3)$.
The CNOTs of $\cx{0}$ and $\cz{0}$ involving $L_0$ would then
both execute in layer 1, creating a resource conflict.
We therefore choose the shared orders and all CNOT execution times
jointly in one CP-SAT model~\cite{perron2023cp}.
Each physical gate retains its own execution time, so corresponding
gates can occupy different layers.

\emph{Shared-order constraints.}
For each check $i\in C$, the shared order $\pi_C$ requires
\begin{equation}
 t_{i,\phi_i(\pi_C(1))}<\cdots<
 t_{i,\phi_i(\pi_C(w_C))}.
 \label{eq:shared-order}
\end{equation}
For each port pair $1\le p<p'\le w_C$, a binary variable
$y_{C,p,p'}$ indicates whether $p$ precedes $p'$.
For every check $i\in C$, we impose
\begin{equation}
 \begin{aligned}
 y_{C,p,p'}=1 &\ \Longrightarrow
 t_{i,\phi_i(p)}+1\le t_{i,\phi_i(p')},\\
 y_{C,p,p'}=0 &\ \Longrightarrow
 t_{i,\phi_i(p')}+1\le t_{i,\phi_i(p)}.
 \end{aligned}
 \label{eq:precedence-link}
\end{equation}
Every feasible assignment to~\eqref{eq:precedence-link} gives each
port pair the same strict precedence in all member checks.
The time inequalities exclude cycles, so these precedences determine
a common permutation $\pi_C$. Conversely, any integer execution times
that follow a common $\pi_C$ satisfy~\eqref{eq:precedence-link} when
the binary variables are set according to that permutation.
The linking constraints therefore encode the shared relative order
exactly, while leaving each CNOT's layer to be chosen under the
resource and measurement constraints.

For the X group in $C_6$, $y_{X,1,2}=1$ places $L_0$ before $L_2$
in $\cx{0}$ and $L_2$ before $L_4$ in $\cx{2}$.
The resource constraint on $L_2$ couples the two checks, while the
measurement constraints couple their times to Z-check gates.
The established resource and measurement constraints~\cite{tremblay2022constant,zhang2026optimal}
act on the time variables in~\eqref{eq:precedence-link}, and therefore
restrict the feasible values of the shared precedence variables.

\begin{figure}[!tbp]
 \centering
\begin{tikzpicture}[x=1.05cm,y=.85cm,font=\small,
 slot/.style={draw=orderblue,fill=orderblue!5,minimum width=.86cm,
              minimum height=.65cm,align=center,inner sep=2pt},
 idle/.style={draw=gray!60,dashed,text=gray,minimum width=.86cm,
              minimum height=.65cm,inner sep=2pt}]
\node[anchor=east] at (.48,1.1) {Layer};
\foreach \xx/\tt in {1/1,2/2,3/3,4/4,5/5}
  \node at (\xx,1.1) {$\tt$};
\node[anchor=east] at (.48,0) {$\cx{0}$};
\node[anchor=east] at (.48,-1.2) {$\cx{2}$};
\foreach \xx/\pp/\qq in {1/1/{L_0},2/3/{R_0},3/4/{R_1},5/2/{L_2}}
  \node[slot] at (\xx,0) {$\pp$\\$\qq$};
\foreach \xx/\pp/\qq in {1/1/{L_2},2/3/{R_2},3/4/{R_3},4/2/{L_4}}
  \node[slot] at (\xx,-1.2) {$\pp$\\$\qq$};
\node[idle] at (4,0) {idle};
\node[idle] at (5,-1.2) {idle};
\node[anchor=west] at (.3,-2.1) {Shared port order: $\pi_X=(1,3,4,2)$};
\end{tikzpicture}

 \caption{\textbf{A shared relative order allows different execution times.}
 Boxes show port labels above data-qubit labels;
 dashed boxes denote idle slots.}
 \label{fig:timing}
\end{figure}
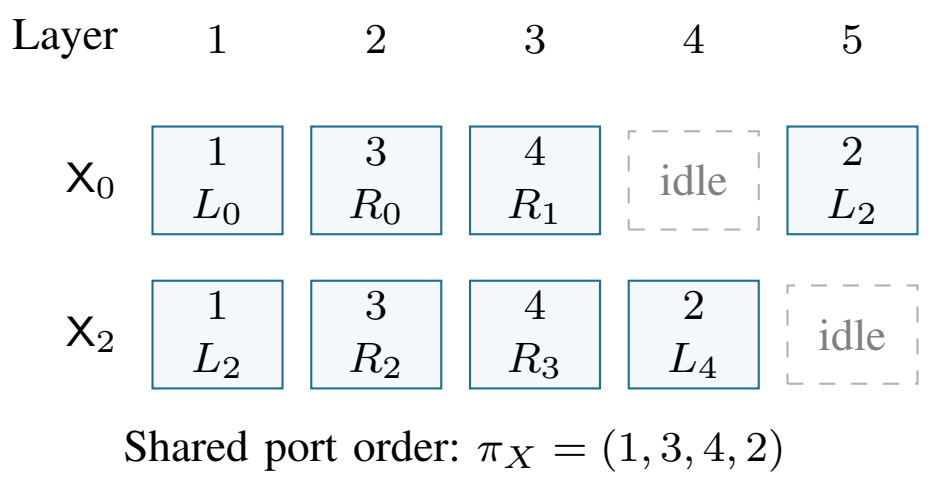

In the $C_6$ example, allowing five layers lets us postpone
the final gate of $\cx{0}$ from layer 4 to layer 5 while
its counterpart in $\cx{2}$ stays at layer 4.
The complete schedule remains valid, and both checks retain
the port order $(1,3,4,2)$ (Fig.~\ref{fig:timing}).
By contrast, when only four layers are available, each weight-four
check must use every layer, so its order fixes its execution times.

\emph{Resource constraints.}
All CNOTs incident to the same ancilla or data qubit
must have distinct execution times.

\emph{Measurement constraints.}
The model also constrains the interleaving of X- and Z-check
CNOTs on shared data qubits.
For each X check $x$ and Z check $z$, we require
\begin{equation}
 \sum_{q\in Q_x\cap Q_z}\ind{t_{x,q}<t_{z,q}}
 \equiv 0\pmod{2},
 \label{eq:interleaving}
\end{equation}
where $\ind{\cdot}$ equals one when its condition is true
and zero otherwise.
The sum counts shared data qubits on which the X-check CNOT
executes before the Z-check CNOT, and this count must be even.

To see why the count must be even, consider exchanging the X-check
and Z-check CNOTs on a shared data qubit. This introduces an additional
CNOT from the X ancilla to the Z ancilla.
These extra CNOTs cancel in pairs when their number is even,
preserving the intended joint measurements%
~\cite{zhang2026optimal}.
In Table~\ref{tab:example}, checks $\cx{0}$ and $\cz{0}$
share $L_0$ and $R_0$.
The X-check CNOT executes first on both qubits, so the sum
in~\eqref{eq:interleaving} is two.
Reversing the precedence on only one shared qubit would make
the sum odd and violate measurement validity.

The shared-order encoding uses
$\sum_{C\in\mathcal{G}}\binom{w_C}{2}$ binary precedence variables.
Applying the same pairwise encoding independently to each check
would require $\sum_i\binom{w_i}{2}$ such variables.
Both representations retain $\sum_i w_i$ CNOT time variables.
For $C_6$, sharing reduces the ordering variables from
$12\binom{4}{2}=72$ to $2\binom{4}{2}=12$, while retaining all
$12\cdot4=48$ time variables. These counts cover the explicit
ordering and time variables. Variables introduced for the other
circuit constraints are additional.

\begin{table*}[!t]
\caption{Performance of circuits synthesized by our method, ASC, PropHunt, and AlphaSyndrome on 15 CSS qLDPC codes}
\label{tab:quality}
\centering
\small
\setlength{\tabcolsep}{4pt}
\renewcommand{\arraystretch}{1.10}
\begin{tabular}{@{}llrrr rrr rrr rrr@{}}
\toprule
& & \multicolumn{3}{c}{Ours}
& \multicolumn{3}{c}{ASC}
& \multicolumn{3}{c}{PropHunt}
& \multicolumn{3}{c}{AlphaSyndrome}\\
\cmidrule(lr){3-5}
\cmidrule(lr){6-8}
\cmidrule(lr){9-11}
\cmidrule(l){12-14}
Code & $\lbrack\!\lbrack n,k,d\rbrack\!\rbrack$
& $D$ & $\dcirc$ & LER
& $D$ & $\dcirc$ & LER
& $D$ & $\dcirc$ & LER
& $D$ & $\dcirc$ & LER\\
\midrule
SB54 & $\lbrack\!\lbrack54,4,8\rbrack\!\rbrack$
& 7 & 7 & \textbf{0.815}
& 7 & 5 & 1.327
& 14 & 4 & 1.643
& 64 & 4 & 7.159\\

TB60 & $\lbrack\!\lbrack60,4,8\rbrack\!\rbrack$
& 7 & 7 & \textbf{0.553}
& 7 & 6 & 0.694
& 18 & 5 & 1.380
& 26 & 4 & 1.993\\

GBB56 & $\lbrack\!\lbrack56,6,8\rbrack\!\rbrack$
& 7 & 6 & \textbf{0.727}
& 7 & 5 & 0.863
& 16 & 5 & 1.140
& 26 & 4 & 1.627\\

BB72 & $\lbrack\!\lbrack72,12,6\rbrack\!\rbrack$
& 7 & 6 & \textbf{1.042}
& 7 & 4 & 1.158
& 14 & 4 & 1.451
& 31 & 4 & 2.146\\

BB144 & $\lbrack\!\lbrack144,12,12\rbrack\!\rbrack$
& 7 & 11 & \textbf{0.033}
& 7 & 8 & 0.038
& 14 & 7 & 0.059
& 44 & 7 & 0.381\\

NA48 & $\lbrack\!\lbrack48,4,8\rbrack\!\rbrack$
& 7 & 6 & \textbf{1.361}
& 6 & 5 & 1.447
& 17 & 4 & 2.138
& 36 & 4 & 3.908\\

NA60 & $\lbrack\!\lbrack60,2,6\rbrack\!\rbrack$
& 8 & 6 & \textbf{1.380}
& 8 & 6 & 1.579
& 18 & 5 & 3.327
& 48 & 4 & 10.128\\

HGP52 & $\lbrack\!\lbrack52,4,4\rbrack\!\rbrack$
& 7 & 4 & \textbf{4.770}
& 7 & 4 & 4.888
& 15 & 4 & 5.901
& 27 & 4 & 7.425\\

HGP66 & $\lbrack\!\lbrack66,1,6\rbrack\!\rbrack$
& 6 & 6 & \textbf{1.948}
& 6 & 6 & 2.044
& 14 & 6 & 2.570
& 25 & 6 & 3.808\\

LP72 & $\lbrack\!\lbrack72,4,8\rbrack\!\rbrack$
& 5 & 7 & \textbf{0.462}
& 5 & 6 & 0.490
& 12 & 6 & 0.564
& 46 & 5 & 2.373\\

LP84 & $\lbrack\!\lbrack84,2,8\rbrack\!\rbrack$
& 5 & 8 & \textbf{0.602}
& 5 & 6 & 0.697
& 9 & 6 & 0.699
& 23 & 6 & 1.228\\

LP135 & $\lbrack\!\lbrack135,31,5\rbrack\!\rbrack$
& 10 & 5 & \textbf{0.952}
& 10 & 5 & 0.985
& 18 & 4 & 1.299
& 45 & 5 & 3.004\\

RQT108 & $\lbrack\!\lbrack108,18,4\rbrack\!\rbrack$
& 7 & 3 & \textbf{3.248}
& 7 & 3 & 3.469
& 16 & 3 & 3.985
& 30 & 2 & 5.369\\

FB126 & $\lbrack\!\lbrack126,8,9\rbrack\!\rbrack$
& 6 & 8 & \textbf{0.155}
& 6 & 7 & 0.175
& 14 & 5 & 0.229
& 40 & 6 & 0.840\\

Haah128 & $\lbrack\!\lbrack128,14,8\rbrack\!\rbrack$
& 8 & 8 & \textbf{0.129}
& 10 & 7 & 0.195
& 19 & 6 & 0.414
& 62 & 5 & 3.102\\
\bottomrule
\end{tabular}

\par\smallskip
\begin{minipage}{\textwidth}
\footnotesize\raggedright
\textit{Note:}
LER is in units of $10^{-3}$; the lowest value in each row is bold.
$D$ denotes CNOT depth; $\dcirc$ denotes circuit-level distance.
\end{minipage}
\end{table*}

\subsection{Circuit Generation and Selection}
\label{sec:selection}

Evaluating~\eqref{eq:objective} requires noisy simulation and decoding.
The joint model ensures measurement validity, while LER evaluation
compares the resulting circuits under noise. Depth counts CNOT
layers, and $\dcirc$ counts the fewest faults causing an undetected
logical error. LER depends on the numbers and
probabilities of harmful fault configurations and the decoder's
response. It can therefore distinguish circuits with equal depth
and distance, and we use it to select among feasible schedules
(Algorithm~\ref{alg:synthesis}).

Depth search starts at the resource lower bound $\Delta$,
the maximum number of required CNOTs incident to any qubit.
We increase the bound within the time budget until the first feasible
schedule is found, then use its depth $D$ for candidate generation.

\begin{algorithm}[!tbp]
 \caption{Structure-guided synthesis and selection}
 \label{alg:synthesis}
 \begin{algorithmic}[1]
 \Require $H_X,H_Z$ and their construction representation
 \Ensure Selected schedule $\sigma^\star$ and its LER

 \State Form sharing groups $\mathcal{G}$ from the supplied check
 correspondences, treating X and Z checks separately
 \State Label a representative check's ports in each group and transfer
 the labels through the supplied index maps to obtain $\{\phi_i\}$
 \State Jointly model shared orders and execution times
 \State From $\Delta$, increase depth until a feasible schedule
 $\sigma$ is found or the time budget is exhausted
 \State \Return failure if no feasible schedule was found
 \State Set $D$ to the depth of $\sigma$;
 $\mathcal{S}\gets\{\sigma\}$

 \While{candidate and time budgets permit another solve}
   \State Solve the joint model within depth bound $D$
   \If{a feasible schedule $\sigma$ is returned and
       $\sigma\notin\mathcal{S}$}
     \State $\mathcal{S}\gets\mathcal{S}\cup\{\sigma\}$
   \EndIf
 \EndWhile

 \State Screen $\mathcal{S}$ by LER and retain a shortlist
 \State Reevaluate the shortlist at higher precision
 \State Select $\sigma^\star$ with the lowest reevaluated LER
 \State \Return $\sigma^\star$ and its independently evaluated LER
 \end{algorithmic}
\end{algorithm}

At depth bound $D$, repeated solves with randomized linear objectives
generate candidates. We remove duplicates by comparing all CNOT
execution times. For example, fixing the $C_6$ Z order to $(4,1,2,3)$
allows the X orders $(1,3,4,2)$ and $(1,4,3,2)$ to produce two valid
four-layer schedules with different fault propagation
(Fig.~\ref{fig:fault}).

\section{Experimental Evaluation}
\label{sec:evaluation}
We evaluate circuit quality, shared-order ablation, time to the first
feasible circuit, and noise sensitivity.
The sharing, timing, and noise experiments use GBB56, TB60, BB72,
Haah128, and BB144 to cover bicycle and lattice constructions
with 56--144 data qubits, including our largest benchmark.

\subsection{Setup and Metrics}
\label{sec:setup}

\emph{Hardware.}
We run the experiments on a system with an AMD EPYC 7K62 CPU,
96 hardware threads, and four NVIDIA RTX PRO 5000 GPUs, each
with 72\,GB of memory.

\emph{Benchmarks.}
Table~\ref{tab:quality} lists 15 code instances from ten families.
These include single-variable bicycle (SB)~\cite{panteleev2021degenerate}, IBM
bivariate bicycle (BB)~\cite{bravyi2024high}, general bivariate
bicycle (GBB)~\cite{lin2024quantum}, non-Abelian two-block group-algebra
(NA)~\cite{lin2024quantum}, and trivariate bicycle (TB)
codes~\cite{voss2025multivariate}. We also evaluate hypergraph
product (HGP)~\cite{tillich2014quantum}, lifted product
(LP)~\cite{panteleev2021quantum}, random quantum Tanner
(RQT)~\cite{leverrier2022quantum}, fiber-bundle
(FB)~\cite{hastings2021fiber}, and Haah codes~\cite{haah2011local}.
These instances test the same synthesis procedure across polynomial,
group-algebra, product, and other repeated-connectivity constructions.

\emph{Baselines and circuit selection.}
We compare with ASC, PropHunt, and AlphaSyndrome
using the same check matrices, memory tasks, and measurement
circuits with one ancilla per check. We use the authors'
implementations~\cite{ascRepository,propHuntRepository,alphaSyndromeRepository}.
PropHunt and AlphaSyndrome retain native search settings, with
AlphaSyndrome receiving feedback from the common Telescoping decoder.
Per code, we run PropHunt ten times. AlphaSyndrome runs for
24 hours and exports one circuit.

For each code, our method and ASC each supply ten candidates.
We screen them by LER, reevaluate the best five per method,
and select one circuit.
For PropHunt, we collect the intermediate and final circuits
from all ten runs, screen the combined pool by LER,
reevaluate the best five candidates, and select one circuit.

\emph{Noise and decoding.}
We evaluate all circuits with the super\-conducting-inspired SI1000
circuit-level noise model~\cite{gidney2021fault}.
Its operation-dependent error strengths approximate the relative
noise costs of gates, preparation, measurement, and idling,
capturing both CNOT fault propagation and idle noise.
The main comparison uses a physical error parameter of $p=0.003$.
Each X- or Z-memory experiment uses $R=d$ rounds of syndrome extraction,
where $d$ is the code distance in Table~\ref{tab:quality}.
Each circuit retains its synthesized CNOT layers and idle periods
during evaluation.

We construct detector error models with
Stim~\cite{gidney2021stim}. All methods use the same Telescoping
decoder configuration~\cite{bhardwaj2026high}, with stages
S1--S3 enabled. For the main comparison, we evaluate final LERs
on independent samples with 5000 decoding failures per memory basis.

\emph{Metrics.}
Using \texttt{dist\_m4ri}~\cite{distM4riRepository}, we compute
circuit-level distances from Stim detector error models with
unit weight per error mechanism. Connected-cluster enumeration gives
lower bounds, whereas randomized-window search gives upper bounds.
These bounds coincide for all reported distances, yielding exact
values for the evaluated detector error models. We report the smaller
of the X- and Z-memory distances as $\dcirc$.
Relative reductions and speedups are calculated from unrounded values.
\subsection{Circuit Quality}
\label{sec:quality}

Across all 15 instances, our selected circuits have lower LER estimates
than ASC, the strongest evaluated baseline in each case.
The median reduction is 11.4\% (Table~\ref{tab:quality}).
The largest LER reductions are 38.6\% relative to ASC,
68.9\% relative to PropHunt, and 95.8\% relative to
AlphaSyndrome. The LER advantage also holds in all thirteen instances
where our circuits and the corresponding ASC circuits have the same depth.

Our circuits have a higher $\dcirc$ than ASC on ten instances
and the same value on five. On seven instances, our circuits
preserve the code distance. For example, compared with
ASC, $\dcirc$ increases from 4 to 6 on BB72 and from 6 to 8 on
LP84. At the same CNOT depth as ASC, our circuits have a higher
circuit-level distance on eight instances and the same distance on five.
On all five instances where both metrics match, our circuits still
have lower estimated LERs. For example, both NA60 circuits have
depth 8 and $\dcirc=6$, yet ours has a 12.6\% lower estimated LER.
On HGP52 and HGP66, all four methods preserve the code distance,
yet our circuits have the lowest LER estimates.

Our circuits also have lower LER on both instances where depths differ. On Haah128, our circuit uses eight layers
instead of ten and has 33.9\% lower LER.
On NA48, by contrast, CP-SAT proves the shared-order model infeasible
at depth six and finds a depth-seven circuit, which still has 6.0\%
lower LER than ASC's depth-six circuit.

\setcounter{topnumber}{2}
\begin{table}[!tbp]
 \centering
 \caption{Effect of shared ordering on circuit depth and LER}
 \label{tab:ablation}
 \footnotesize
 \setlength{\tabcolsep}{5pt}
 \renewcommand{\arraystretch}{1.08}
 \begin{tabular}{@{}lcrr@{}}
 \toprule
 & Depth & \multicolumn{2}{c}{LER ($\times10^{-3}$)}
 \\
 \cmidrule(l){3-4}
 Code & S/N & Shared & No sharing\\
 \midrule
 GBB56 & 7/7
 & $0.743\pm0.038$ & $0.868\pm0.029$\\
 TB60 & 7/7
 & $0.551\pm0.010$ & $0.636\pm0.035$\\
 BB72 & 7/8
 & $1.039\pm0.015$ & $1.241\pm0.020$\\
 Haah128 & 8/12
 & $0.129\pm0.005$ & $0.241\pm0.004$\\
 BB144 & 7/8
 & $0.0312\pm0.0002$ & $0.0449\pm0.0002$\\
 \bottomrule
 \end{tabular}
 \par\smallskip
 \begin{minipage}{\columnwidth}
 \scriptsize\raggedright
 Depth pairs list shared/no-sharing (S/N) variants.
 LER values are mean $\pm$ SD over three batches.
 \end{minipage}
\end{table}

\subsection{Effect of Shared Ordering}
\label{sec:ablation}

To examine the effect of shared ordering, we enable or disable
the shared-order constraints in the same synthesis implementation.
We keep the remaining constraint encoding, solver settings, candidate
budgets, and selection procedure fixed.
Without sharing, each check's gate order varies independently,
whereas enabling sharing gives checks with corresponding connectivity
a common relative order.
For each code in Table~\ref{tab:ablation}, we run three
batches of candidate generation and selection per variant.
Each batch yields one selected circuit, and Table~\ref{tab:ablation}
reports the mean and standard deviation across the three circuits
obtained for each variant.

Shared ordering reduces the mean LER by 13.4\%--46.6\% across
the five codes and gives a lower LER in all fifteen paired
batch comparisons. GBB56 and TB60 have equal depths, with LER
reductions of 14.4\% and 13.4\%, respectively.
For the remaining three codes, shared ordering yields both shallower
circuits and lower LER. In these cases, the solver without sharing
finds no feasible circuit at the shared variant's depth within the
allotted time and therefore proceeds to greater depths.
Thus, shared ordering improves the quality delivered by this synthesis
procedure, both at unchanged depth and when it helps the solver find
shallower circuits within the time budget.

\raggedbottom
\flushcolsend
\begin{figure}[!tbp]
 \centering
 \includegraphics[width=\columnwidth]{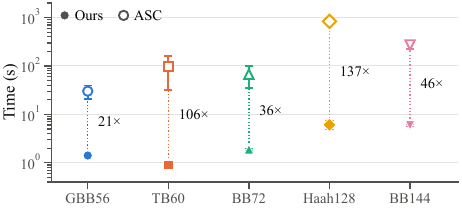}
 \caption{\textbf{Time to the first feasible circuit.}
 Filled and open markers show our method and ASC, respectively.
 Markers and error bars indicate the mean and standard deviation
 over five runs. Annotations show speedups relative to ASC.}
 \label{fig:runtime}
\end{figure}

\subsection{Time to the First Feasible Circuit}
\label{sec:runtime}

Our method reduces the mean time to the first feasible circuit
by factors of 21.4--136.6 relative to ASC across all five timed codes.
Among the baselines, we time ASC because it also builds circuits
with a constraint solver, whereas PropHunt and AlphaSyndrome run
multi-hour iterative searches in our experiments.
We time modeling and solving from the resource depth lower bound
to the first feasible circuit, excluding LER evaluation.
Both methods use eight threads on the same CPU and five seeds per code,
with a 300-second solver limit per depth.

On GBB56, TB60, and BB72, both methods return depth-seven
circuits, but ours is 21.4, 106.1, and 35.6 times faster,
respectively (Fig.~\ref{fig:runtime}).
On Haah128 and BB144, it also reaches equal or lower depths,
running 136.6 and 45.9 times faster,
returning depths 8 and 7 versus ASC's 10--11 and 7--8,
respectively.

\begin{figure}[!tbp]
 \centering
 \includegraphics[width=\columnwidth]{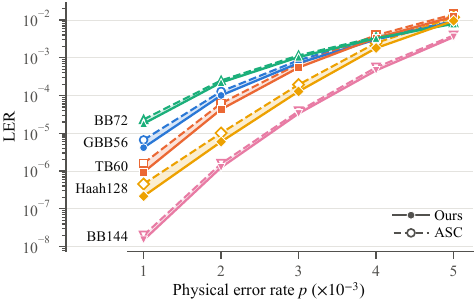}
 \caption{\textbf{Noise sensitivity of the selected circuits.}
 Colors and markers identify codes.
 Solid lines with filled markers show our method;
 dashed lines with open markers show ASC.
 Shading marks the gap between paired curves.}
 \label{fig:noise}
\end{figure}

\subsection{Noise Sensitivity}
\label{sec:noise}
Circuits selected at $p=0.003$ retain their LER advantage over ASC
when evaluated at other physical error rates.
To assess this behavior, we evaluate both methods' selected circuits
from Table~\ref{tab:quality} on the five codes used above, scanning
five physical error rates from $p=0.001$ to $0.005$ while keeping
gate orders, execution layers, and all other noise and decoding
settings fixed. Our circuits have lower LER estimates at all 25 points
(Fig.~\ref{fig:noise}).
The reductions relative to ASC range from 22.8\%--52.4\% at
$p=0.001$ to 5.1\%--19.2\% at $p=0.005$, showing a larger
advantage at lower noise.

\section{Conclusion}
Our method uses construction relationships to organize shared gate
orders and jointly determine them with individual CNOT execution times
under resource and measurement constraints.
The resulting circuits have the lowest LER estimates among the
evaluated methods on all 15 structured CSS qLDPC codes.
The advantage holds in all thirteen equal-depth comparisons with ASC.
They also improve circuit-level distance over ASC on ten instances
and preserve code distance in seven.
Within the same synthesis implementation, sharing improves the quality
of the selected circuits. The complete method finds its first feasible
circuit faster than ASC at equal or lower depth.
The selected circuits retain lower LER estimates across the tested
noise range.
These results show that code structure can guide the synthesis of
lower-error syndrome-extraction circuits.

The source code is available upon request.

\begin{samepage}
\section*{AI Assistance}
Large language models assisted only with language polishing. The authors reviewed and edited all assisted text, provided all scientific ideas, experiments, and results, and take full responsibility for the manuscript.

\end{samepage}

\clearpage
\raggedcolsend
\raggedend
\twocolumn
\raggedbottom
\IEEEtriggeratref{14}
\bibliographystyle{IEEEtran}
\bibliography{ref}
\end{document}